\documentclass[galaxies,article,subbmit,moreauthors,pdftex,10pt,a4paper]{mdpi}

\usepackage{graphicx}
\usepackage{natbib}
\usepackage{pdflscape} 
\usepackage{amssymb}
\usepackage{ulem} 
\usepackage{color} 

\newcommand{\be}{\begin{equation}}
\newcommand{\ee}{\end{equation}}

\newcommand{\ba}{\begin{eqnarray}}
\newcommand{\ea}{\end{eqnarray}}

\newcommand\eg{{\it{{e.g.,\ }}}}

\newcommand{\Lf}{{Lorentz factor}}
\newcommand{\Bf}{{magnetic field}}

\newcommand{\apj}{ApJ}

\newcommand{\aj}{AJ}
\newcommand{\mnras}{MNRAS}
\newcommand{\nat}{Nature}

\newcommand{\araa}{ARA\&A}

\newcommand{\aap}{A\&A}

\newcommand{\nar}{{NAR}}

\externaleditor{Academic Editor: Jose L. Gomez, Alan P. Marscher, Svetlana G. Jorstad}
\history{Received: date; Accepted: date; Published: date}

\Title{Emission knots and polarization swings of swinging jets}

\Author{Maxim Lyutikov $^{1,}$* and  Evgeniya Kravchenko $^{2}$}
\AuthorNames{Maxim Lyutikov , Evgeniya Kravchenko}

\address{%
$^{1}$ \quad Department of Physics, Purdue University, 
 525 Northwestern Avenue,
West Lafayette, IN
47907-2036, USA; lyutikov@purdue.edu\\
$^{2}$ \quad Lebedev Physical Institute, Astro Space Center, Profsoyuznaya 84/32, Moscow 117997, Russia; evgenia.v.kravchenko@gmail.com}

\corres{Correspondence: lyutikov@purdue.edu}

\abstract{Knots (emission features  in  jets of active galactic nuclei)  often show non-ballistic dynamics and variable emission/polarization properties.  We model these features as emission pattern propagating in a jet that carries  helical \Bf\ and is launched along   a changing  direction. The model can reproduce a wide range of phenomena  observed in the motion of knots:  non-ballistic motion (both smooth and occasional  sudden change of direction, and/or oscillatory behavior),  variable brightness, confinement of knots' motion within an overlaying envelope. The model also reproduces  smooth large polarization angle swings, and at the same time allows for the seemingly random  behavior of synchrotron fluxes, polarization fraction  and occasional $\pi/2$  polarization   jumps.}


\begin{document}



\section{Introduction}

Blazars --  a sub-class of active galactic nuclei (AGN) -- have the orientation of their jets close to the line of sight (LoS). This causes their non-thermal radiation to be highly relativistically beamed. 
Their linear fractional polarization reaches values up to 50 per cent \citep[\eg][]{2005AJ....130.1389L} suggesting  the presence of highly ordered  magnetic fields in their compact regions \citep{2005MNRAS.360..869L}.
{Furthermore, the observed behavior of the polarization degree and angle suggest a helical shape of these magnetic fields} \citep[\eg][]{1999NewAR..43..691G,2005MNRAS.356..859P,2005MNRAS.360..869L}.

Blazars are observed to show high variability across the electromagnetic spectrum \citep[\eg][]{1989Natur.337..442Q,1997ARAA..35..445U}. 
The evolution of the $\gamma$-ray, optical, radio and polarized fluxes often exhibit seemingly random behavior \citep[\eg][and references therein]{2013ApJ...768...40L} and in some cases might be represented by a number of isolated, individual events superimposed on a steady processes \citep[\eg][]{2008Natur.452..966M,2016AA...590A..10K}.
In contrast, the optical electric vector position angle (EVPA) variations often show smooth swings of the linearly polarized radiation, with total rotations up to a few radians. Apparent similarities of optical flux, degree of polarization and EVPA during these events, detected in  different sources and their different flaring states, suggests a common mechanism being responsible for  such  a behavior.
The general pattern includes high variability of polarization $\Pi$, which drops to zero during the middle of an EVPA swing and then recovers back to the initial value, smooth and continuous change of polarization angle and peaked behavior of optical flux density.

\section{Polarization swings: jet with helical \Bf\ propagating along a variable direction}

\begin{figure}[h!]
\includegraphics[width=.95\columnwidth]{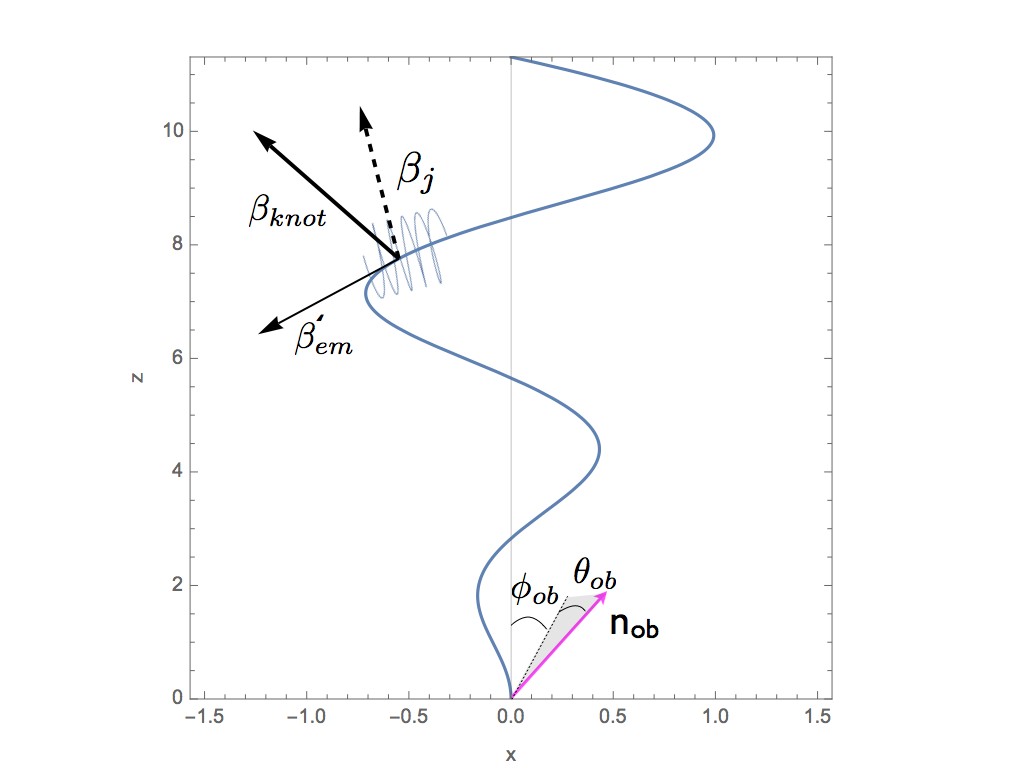}
\caption{Schematic representation of the model. The jet is emitted along a variable direction (defined, \eg\ by the opening angle of the planar motion, jet's oscillation angle). Solid line is a snapshot of the jet at a given moment. Pictured is  the shape of the jet  at some moment $t$. The internal  helical structure of the \Bf\ within the jet is aligned with the local jet direction and changes with time. Physical motion of jet's fluid is with velocity $\beta_j$ along a ballistic trajectory (dashed arrow). In addition, a feature propagates along the instantaneous direction of the jet with velocity $\beta_{em}'$ (thin arrow). We assume that the feature propagates toward the core (the origin) in the jet frame.   The combined ballistic  motion of the fluid and of the pattern along the jet results in the non-ballistic trajectory of a pattern.
Total velocity of the knot is $\beta_{knot}$ - it is non-ballistic. (Velocity vectors do not form a closed triangle due to non-linearity of relativistic velocity addition.) Direction to the observer ${\bf n}_{ob}$ makes angle $\theta_{ob}$ with the plane of the  jet's motions; projection of  ${\bf n}_{ob}$ onto the plane of the jet's motion (thin dashed line) makes an angle $\phi_{ob}$ with respect to the symmetry axis.} 
\label{Picture-Main}
\end{figure}

We present a model (Lyutikov \& Kravchenko 2016a), shown in Fig. \ref{Picture-Main} - a jet propagating along a smoothly variable direction carrying helical \Bf\  - which is able to reproduce large smooth variations of the EVPA, yet allow for occasional sudden jumps in EVPA. In addition - and most importantly -  the intensity and  polarization fraction, though produced by a highly deterministic process, show large non-monotonic variations that can be mistaken for a random process. Thus, a highly deterministic set-up of  the model  produces both smooth variation of EVPA and yet allows for some properties of the emission to vary in a non-monotonic way, which can be interpreted as stochastic variation.

We  model the  emitting element as a jet carrying helical  \Bf\, with internal  pitch angle $\psi$, propagating with \Lf\  $\gamma_j$. The jet produces polarized  synchrotron emission.  
We concentrate on the optically thin region, sufficiently far downstream of the core. In terms of physical location the model is applicable to  sub-parsec to parsec scale regions of the jet.  In the present paper we do not make a separation between the different parts of the spectrum, \eg\ optical and radio, but outline the general properties of polarized synchrotron emission expected from a jet with variable direction.

Calculations of  polarization produced by relativistically moving sources is somewhat complex \citep{1979ApJ...232...34B,2003ApJ...597..998L,2005MNRAS.360..869L}. 
Conventionally (and erroneously for a relativistically moving plasma!), the direction of the observed polarization for optically thin regions and the associated magnetic fields are assumed to be in one-to-one correspondence, being orthogonal to each other, so that some observers choose to plot the direction of the electric vector of the wave, while others plot vectors orthogonal to the electric vectors and call them the direction of the magnetic field. 
{\it This  is  correct only for non-relativistically  moving optically-thin sources, and thus cannot be applied to AGN jets.}
Since the emission is boosted by the relativistic motion of the jet material, {\it  the  EVPA rotates parallel to the plane containing the line of sight and the plasma velocity vector}, so that {\it the observed electric field of the wave is not, in general, orthogonal to the observed magnetic field},  \citep{2003ApJ...597..998L,2005MNRAS.360..869L}. 

In case of unresolved jets the average polarization for an axially symmetric jet can be only along or perpendicular to the jet direction. Yet the same jet can produce the average polarization along or perpendicular depending on its \Lf\ and the viewing angle.

We  consider the synchrotron  emission of  an unresolved, thin, circular  cylindrical shell populated by relativistic electrons with a power law distribution and moving uniformly in the axial direction with constant velocity.  
The properties of the synchrotron emission are then determined by {\it three parameters}: the internal pitch angle of the magnetic field $\psi$, Lorentz factor of the shell in the laboratory frame  $\gamma_j$ and the viewing angle, $\theta$, which the line of sight to the observer makes with the jet axis in the observer reference frame.
Thus, even for fixed internal parameters of the jet, the resulting polarization signature strongly depends both  on the viewing angle and the jet \Lf.

The polarization direction from an unresolved jet can be either along the projection of the jet onto the plane of the sky, or perpendicular to it. Thus, as a jet's direction changes with time, the direction of polarization will also change. Most of the time, the EVPA will either be always along or across the jet. In addition, for a fairly narrow range of internal pitch angles and lines of sight a given jet can show ninety degree EVPA flips.

\begin{figure}[h!]
\centering 
$\theta_{ob,0}=1/(5 \gamma), \hskip .15 \columnwidth  \theta_{ob,0}=1/\gamma,  \hskip .15 \columnwidth  \theta_{ob,0}=\, 2/\gamma$
\\
\includegraphics[width=.30\columnwidth]{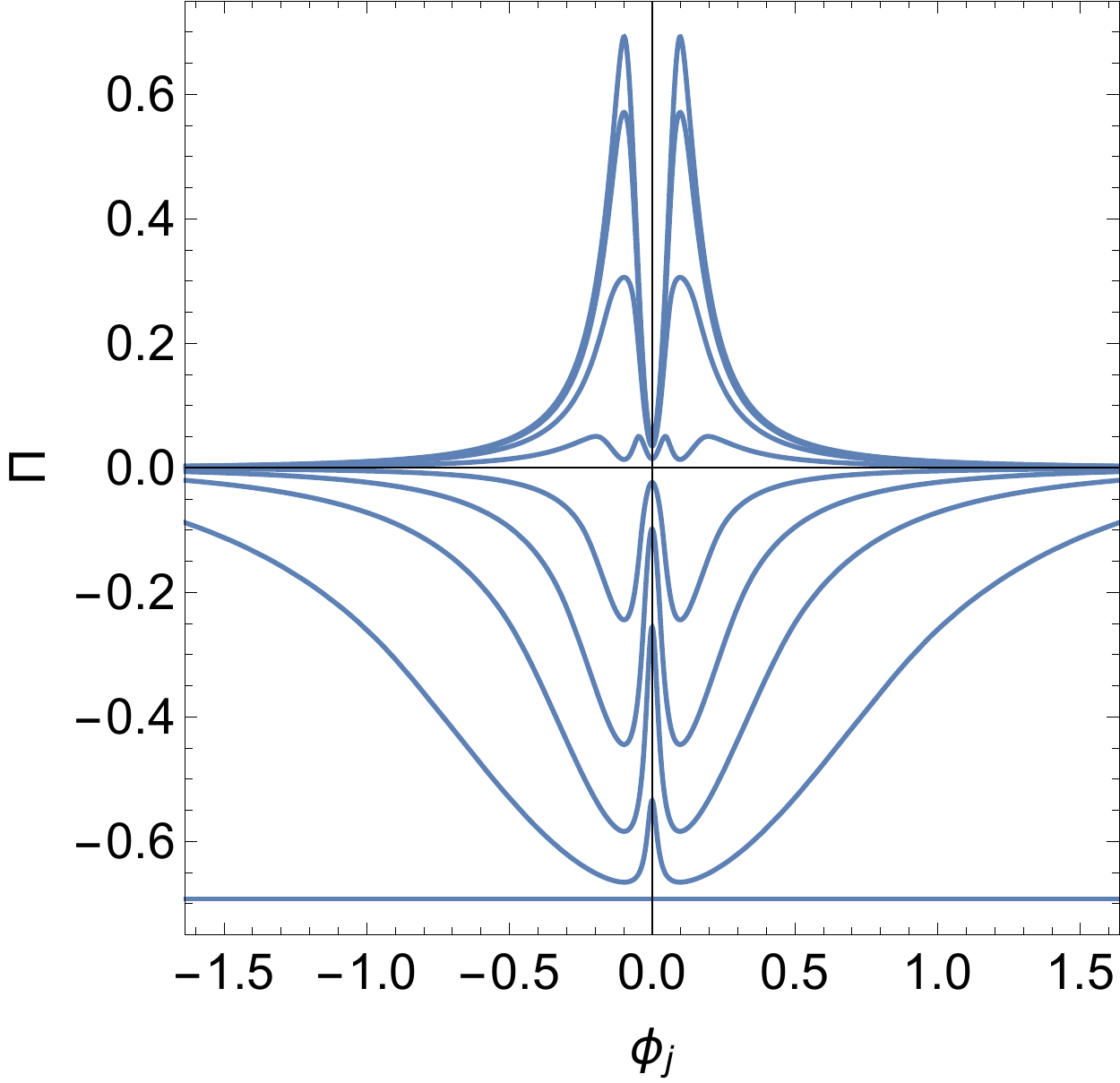}
\includegraphics[width=.30\columnwidth]{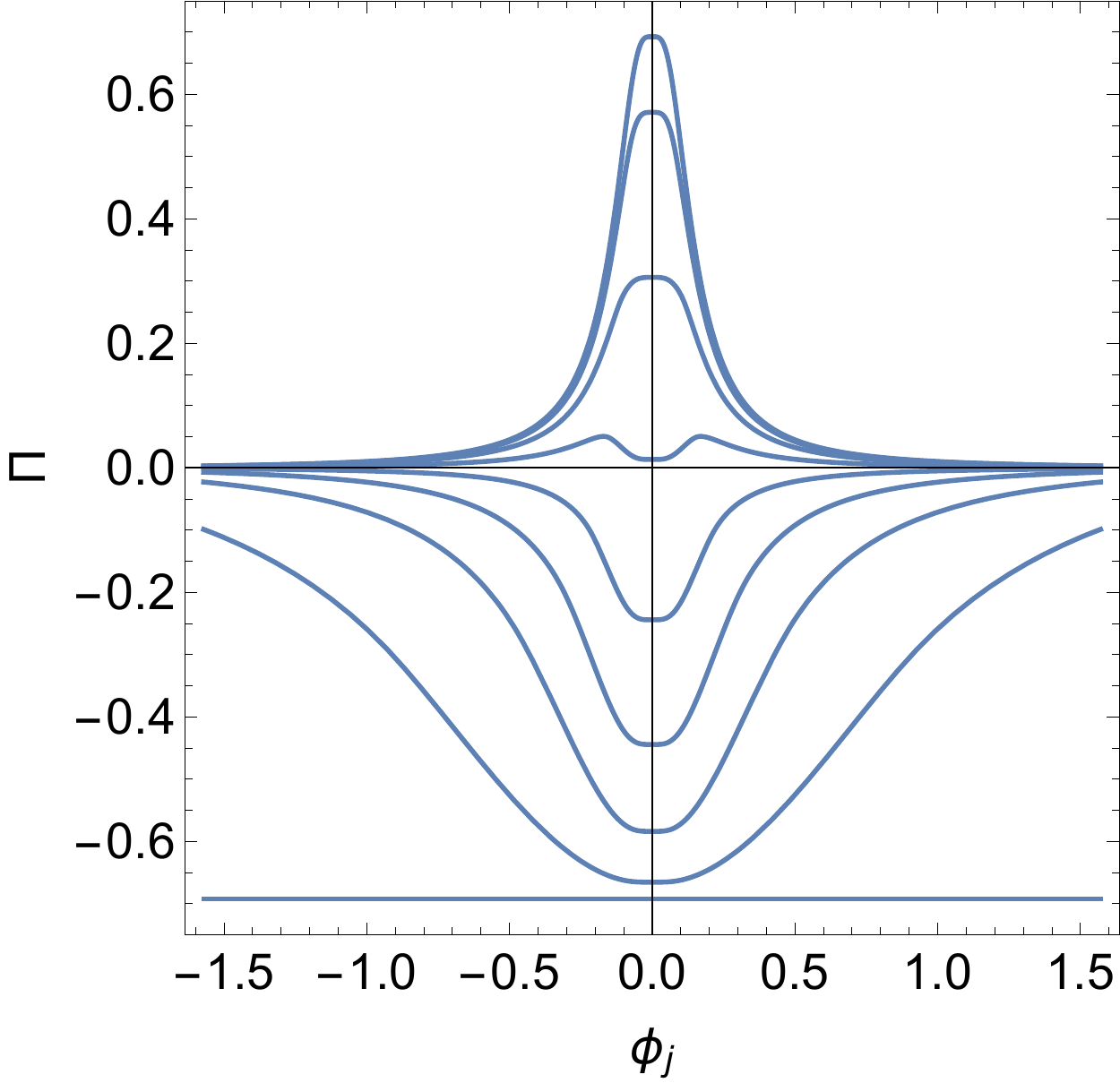}
\includegraphics[width=.30\columnwidth]{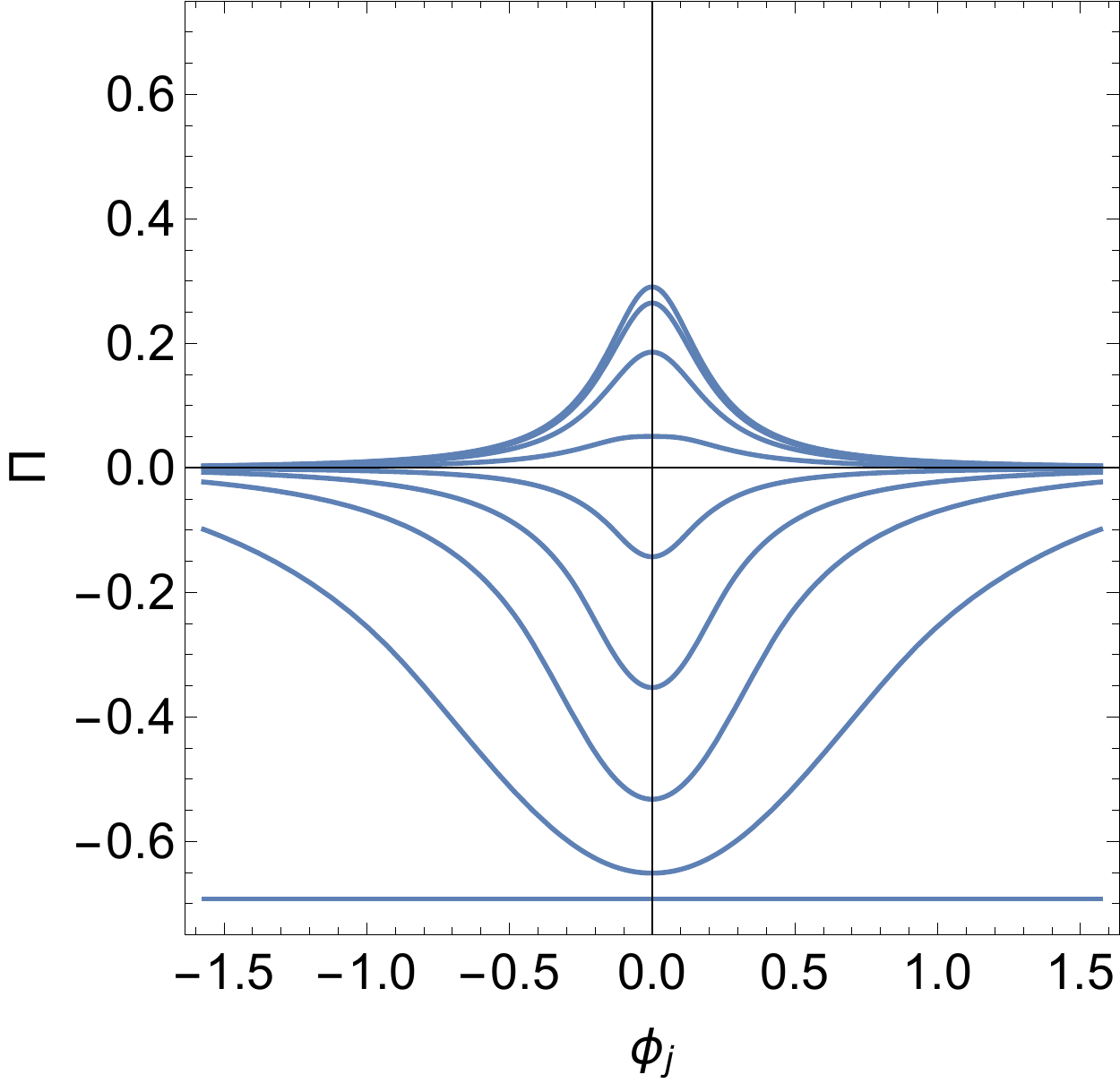}
\\
\includegraphics[width=.30\columnwidth]{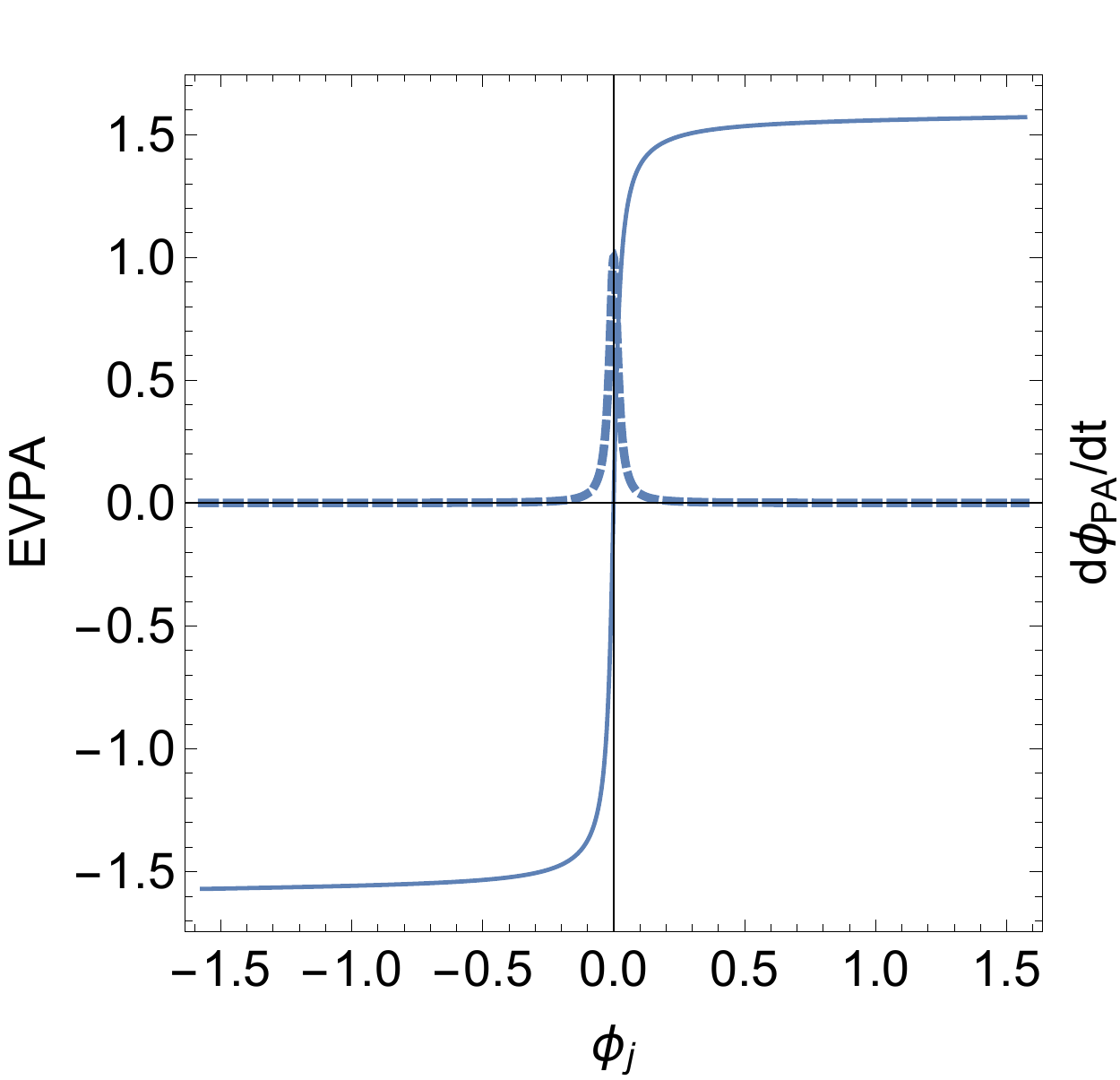}
\includegraphics[width=.30\columnwidth]{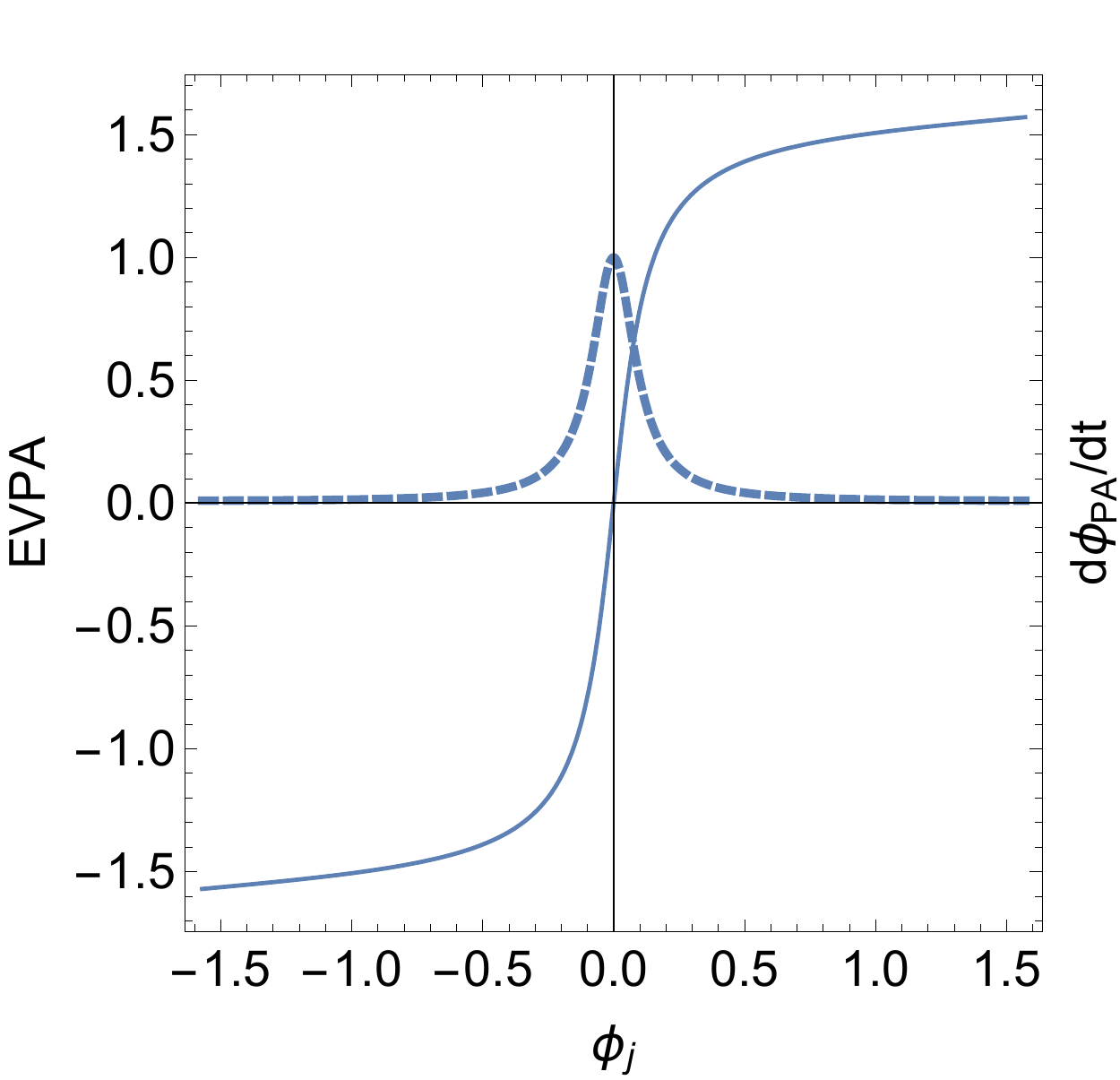}
\includegraphics[width=.30\columnwidth]{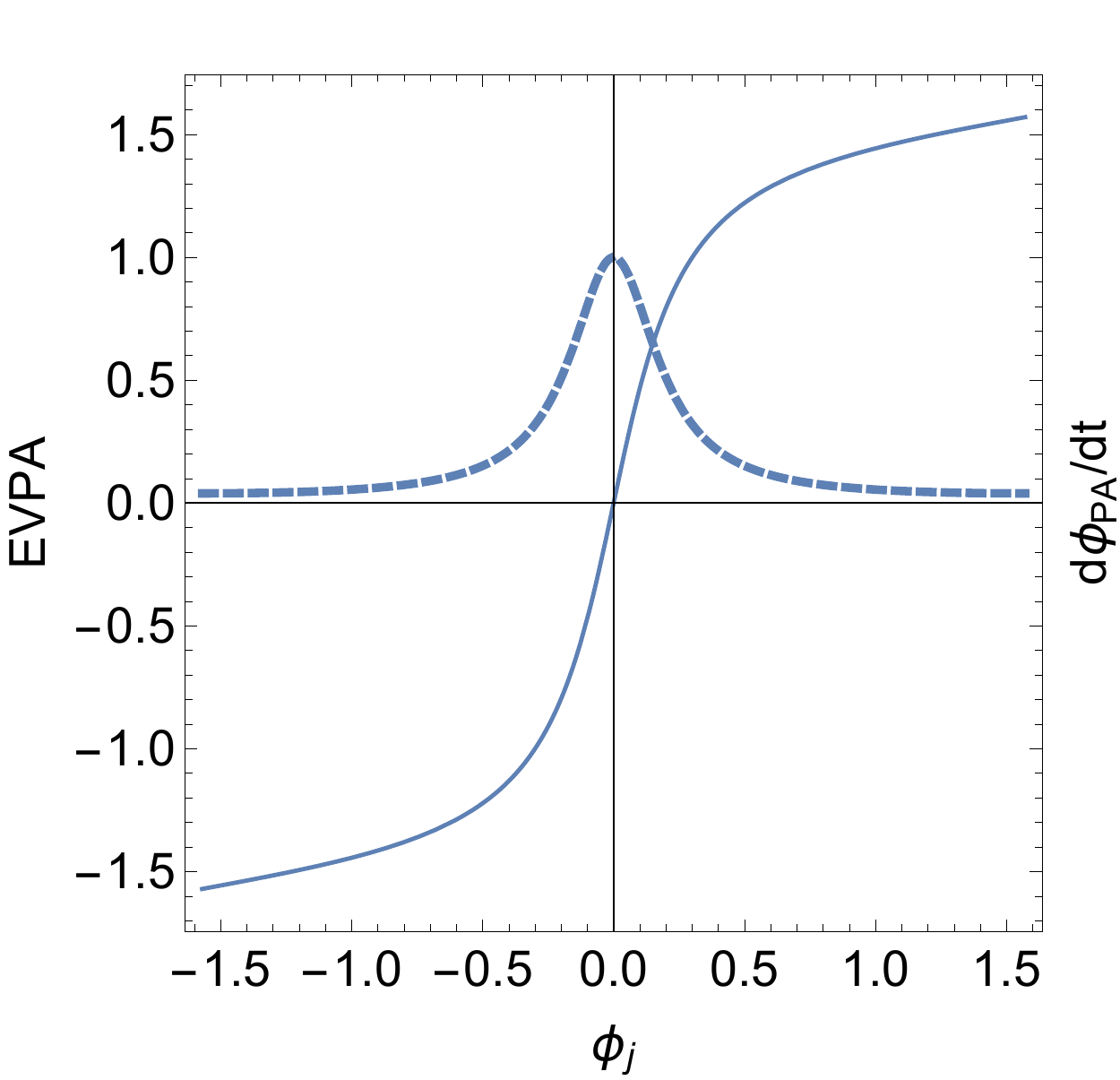}
\caption{Polarization $\Pi$ and EVPA  for a jet executing planar motion. The jet is moving with bulk \Lf\  $\gamma=10$ and is viewed at the minimal viewing angles of  $\theta_{ob,0}=1/(5 \gamma), \, 1/\gamma, \, 2/\gamma$ (left to right columns). {\it Top Row}: $\Pi$ as function of the oscillation angle  for different  intrinsic pitch angles (pitch angels are $0, \pi/16/\pi/8 ... \pi/2$).  {\it Bottom Row}:  EVPA as function of the oscillation angle (solid line). Here a larger range of angles $\phi_j$ is plotted to show the full periodic behavior of EVPA.  Dashed line: the rate of change of EVPA, $\dot{\phi}_{PA}$ (defined here as the projection of the jet on the plane of the sky - EVPA may differ by $90^\circ$), normalized to the maximal value.
} 
\label{Piofphij}
\end{figure}

\begin{figure}[h!]
\centering 
$\theta_{ob,0}=1/(5 \gamma), \hskip .15 \columnwidth  \theta_{ob,0}=1/\gamma,  \hskip .15 \columnwidth  \theta_{ob,0}=\, 2/\gamma$
\\
\includegraphics[width=.30\columnwidth]{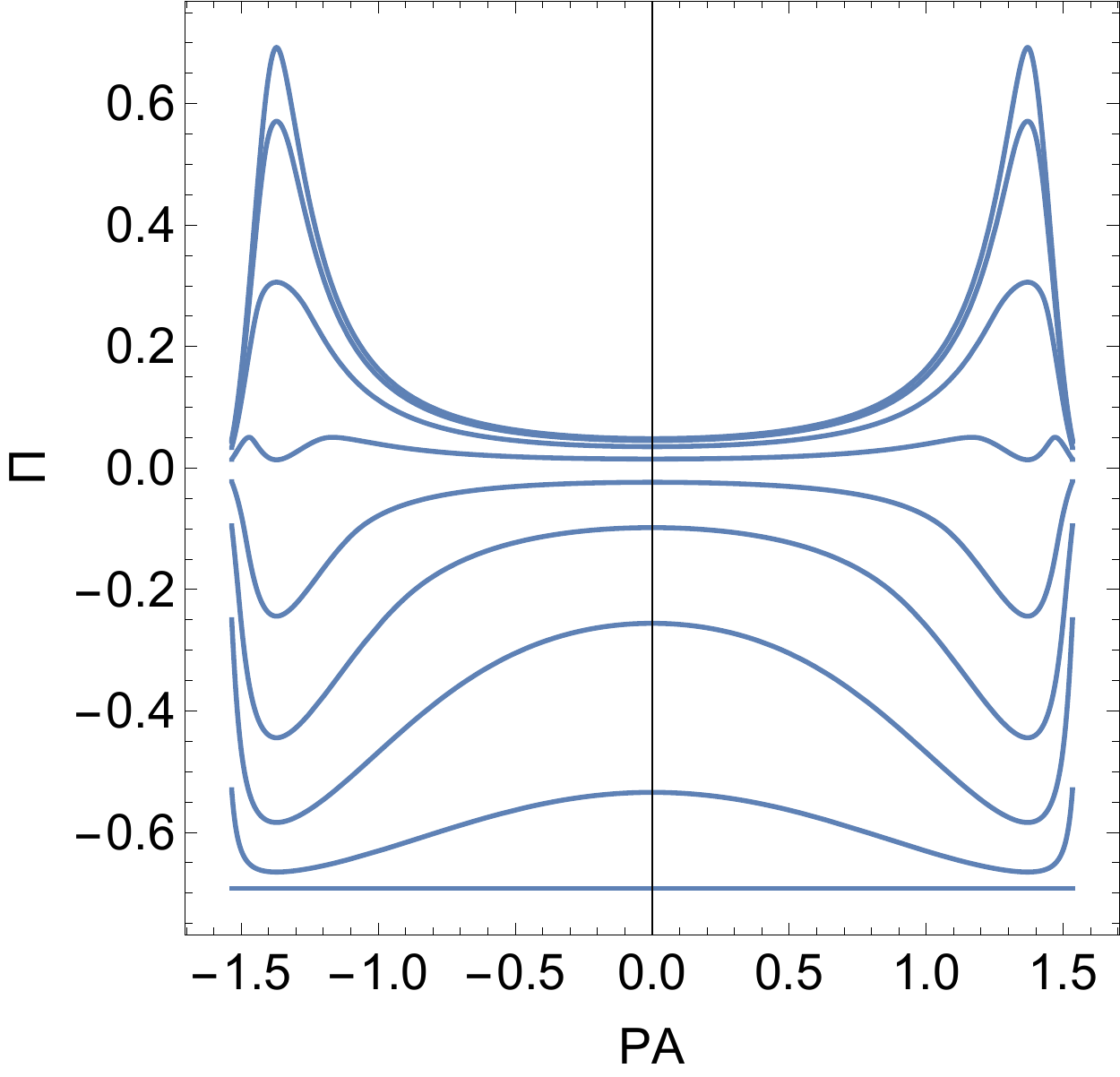}
\includegraphics[width=.30\columnwidth]{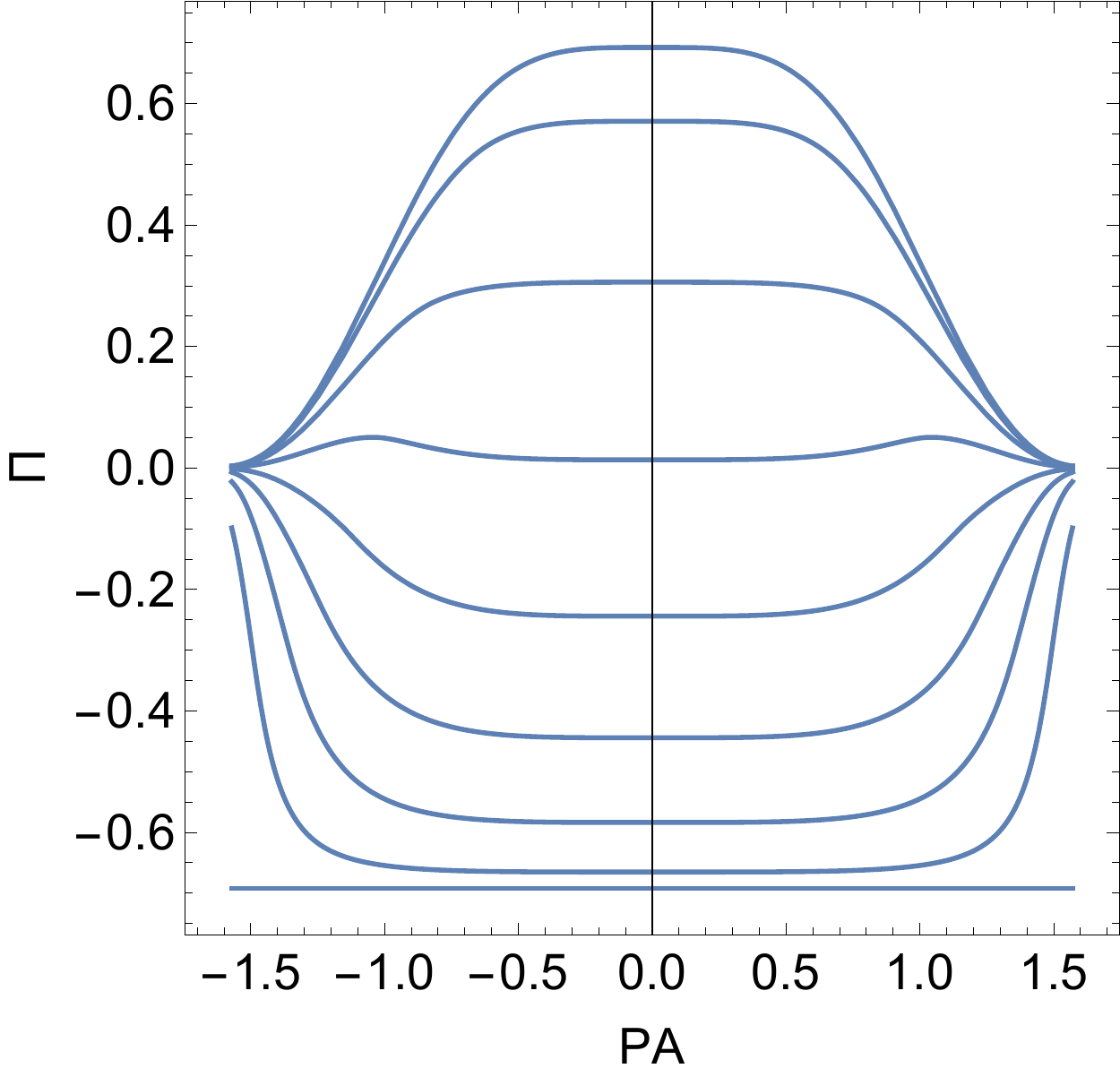}
\includegraphics[width=.30\columnwidth]{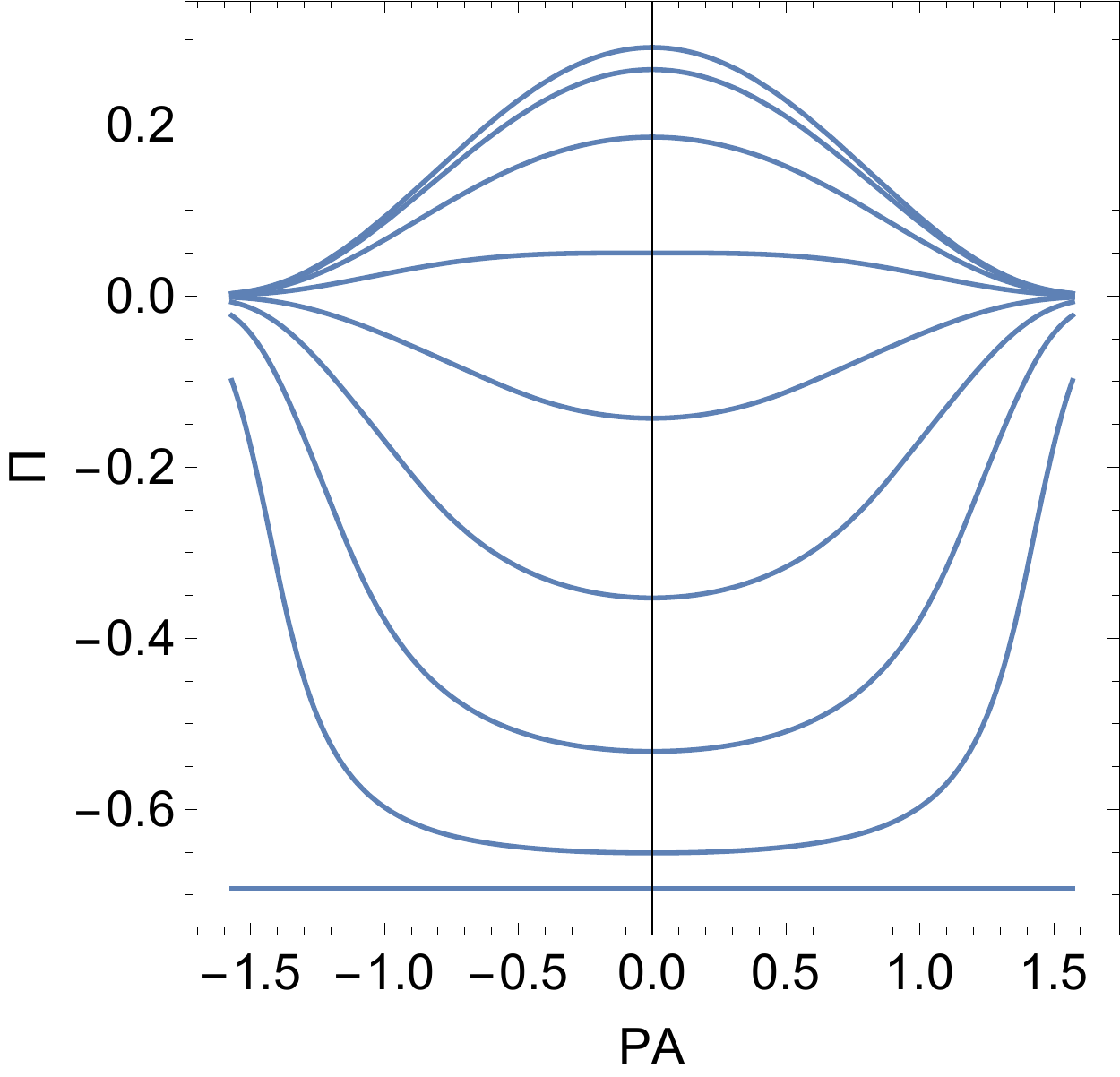}
\\
\includegraphics[width=.30\columnwidth]{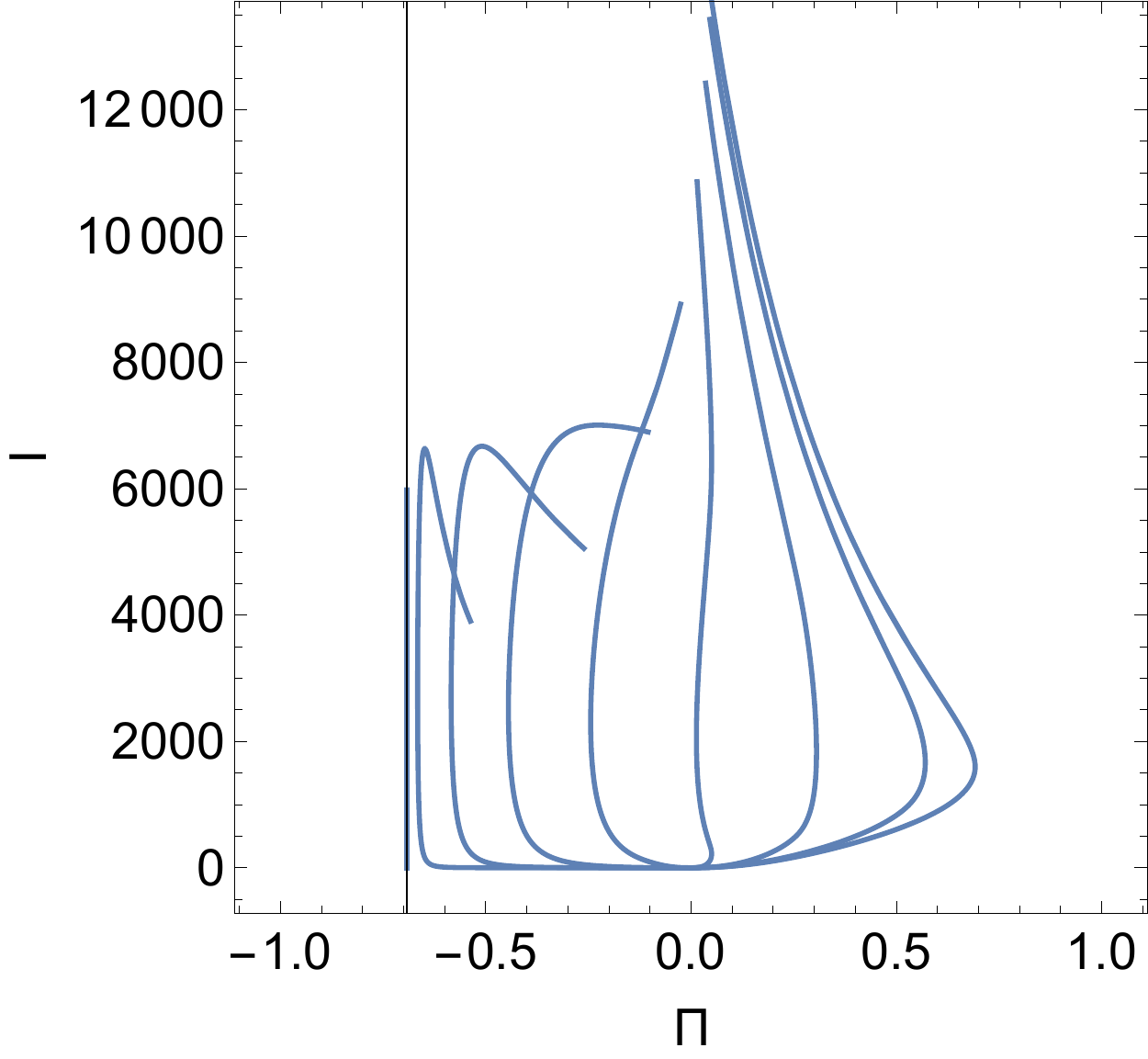}
\includegraphics[width=.30\columnwidth]{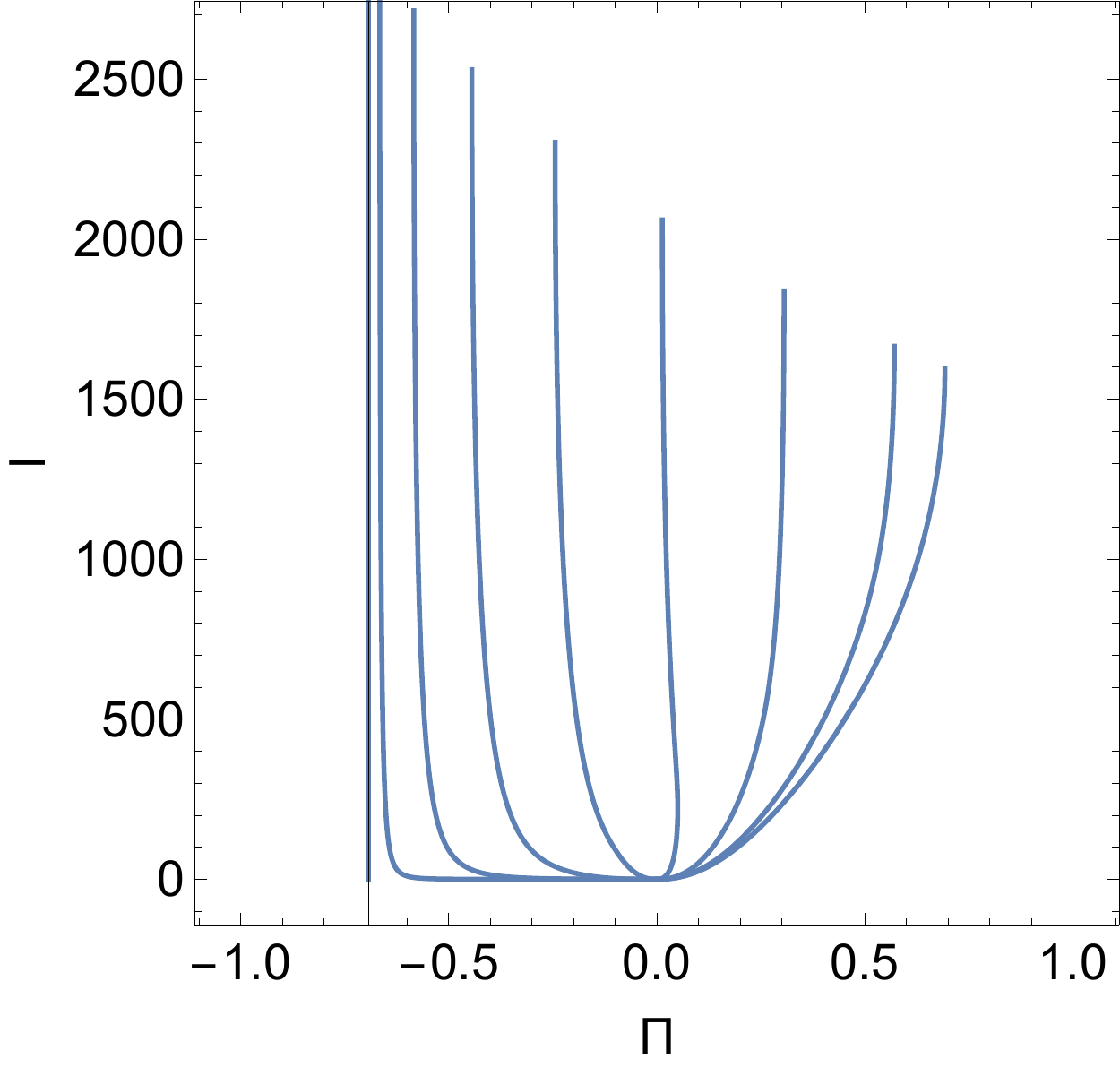}
\includegraphics[width=.30\columnwidth]{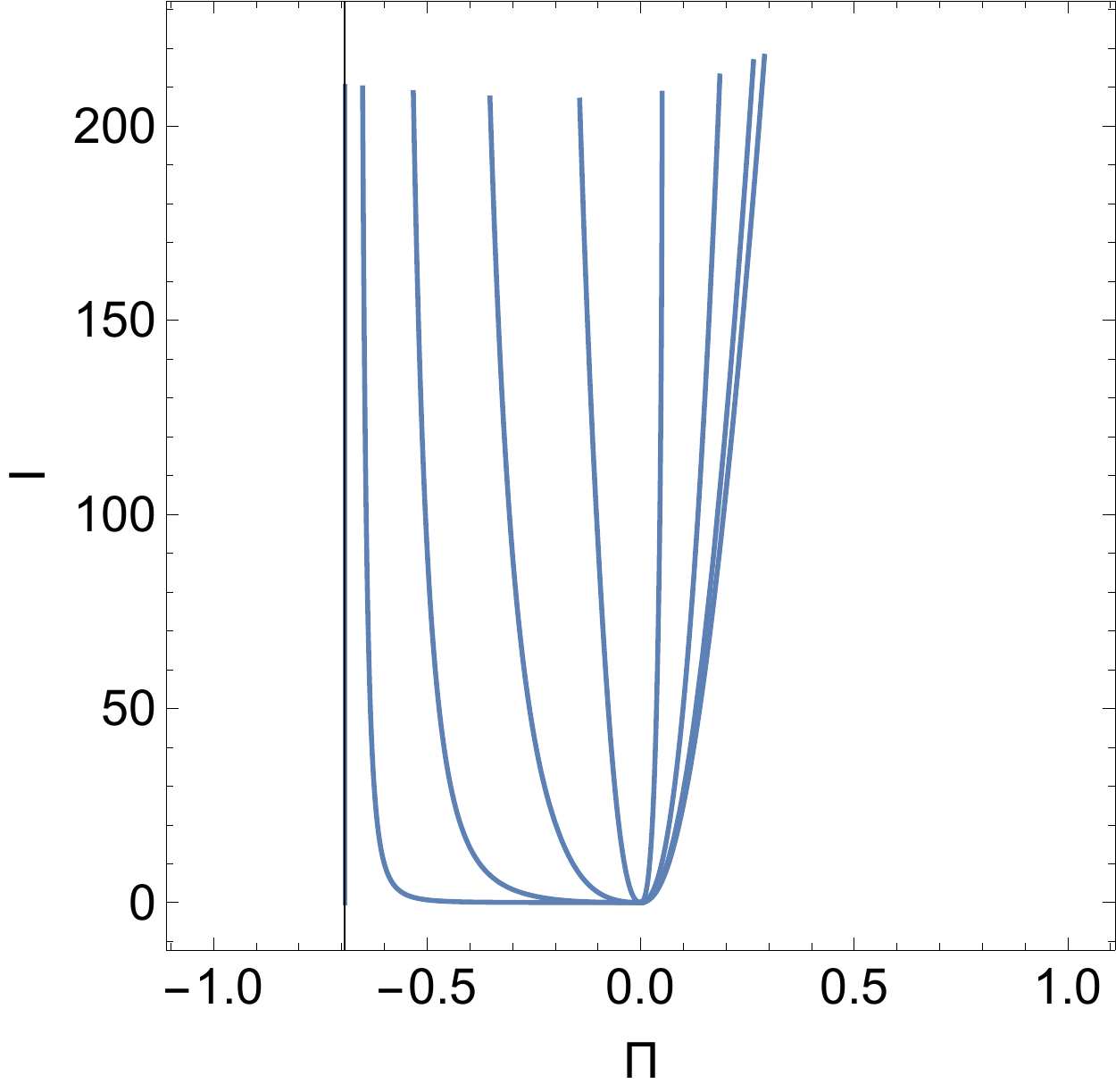}
\caption{Same set-up as Fig. \protect\ref{Piofphij}. {\it Top Row}: polarization  $\Pi$ as function of position angle. {\it Bottom Row}: Intensity as function of  polarization degree $\Pi$).
} 
\label{Piofphij10}
\end{figure}

In Fig. \ref{Piofphij}-\ref{Piofphij10} we plot the  polarization signatures assuming that the motion of the jet is symmetric with respect to the line of sight and that oscillations occur between angles $-\pi/2 < \phi_j < \pi/2$ ($\phi_j$ is the angle between the instantaneous jet direction and the symmetry axis).
{We note, for $\Pi>0$ the polarization is along the jet, while the polarization is orthogonal to the jet for $\Pi<0$.}

\subsection{Overall trends and special cases}


\begin{itemize}
\item Fastest swings of EVPA typically coincide with intensity and polarization extrema (maxima or minima).

\item  EVPA can experience fast $\pm \pi/2 $ jumps near $\Pi=0$.
At these points intensity can be maximal or minimal.

\item Swinging jets can produce very fast EVPA swings, up to  $\pm \pi$; at the maximum rate of the EVPA swings (second   rows in Fig. \ref{Piofphij})  the   polarization fraction minima can be maximal or minimal (first rows in Fig. \ref{Piofphij}).


\item Intensity depends on the  EVPA and the polarization fraction in a complicated way; for accelerating jets such  dependence is not necessarily symmetric.

\item {Swinging jets can give EVPA rotations in opposite directions within the same source (Fig. \ref{Piofphij}, change in oscillating angle from $\pi$/2 back to $-\pi$/2).}

\item {Circularly moving and swinging jets can produce EVPA rotations with  amplitude $\leq\pi/2$, including complex changes 
 if a jet experiences only slight changes in its direction.}

\end{itemize}

We stress that these very complicated,  apparently random, and yet correlated variations of intensity, polarization fraction and EVPA swings come from the assumed highly regular jet motion and very regular jet structure.
Most importantly, we assumed {\it constant}  jet emissivity. We expect that variations in the acceleration/emissivity properties of the jets will further complicate the observed properties. 
As well as more complicated trajectories of a beam can give more complicated profiles.

\section{Knots as emission patterns in swinging AGN jets}

Blazar jet often show emission features that are moving along non-ballistic trajectories \citep{2016AJ....152...12L}. 
Lyutikov \& Kravchenko (in preparation) discuss  a kinematic model of the emission features as disturbances propagating along a  ballistically moving jet.

Let a  jet moving ballistically  with velocity $\beta_j$  oscillate in the $x - z$ plane. In addition, let an emitting element propagate along the jet with velocity $\beta_{em}'$ with respect to the jet. In the observer frame the velocity of the element is  given by
relativistic addition of the radial fluid outflow and velocity along the jet, Fig. \ref{Picture-Main}.

\begin{figure}[h!]
\includegraphics[width=.45\columnwidth]{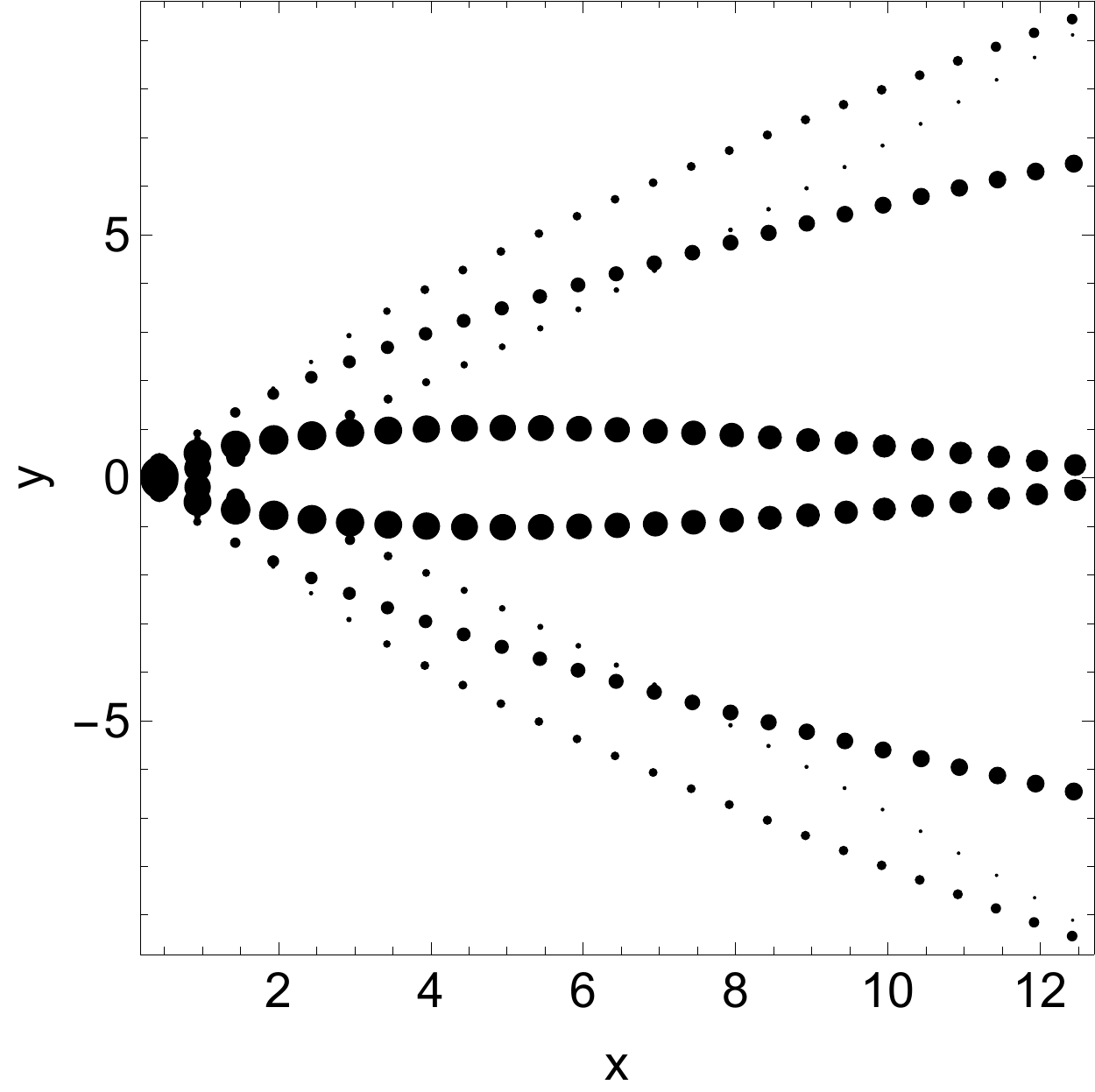}\quad
\includegraphics[width=.45\columnwidth]{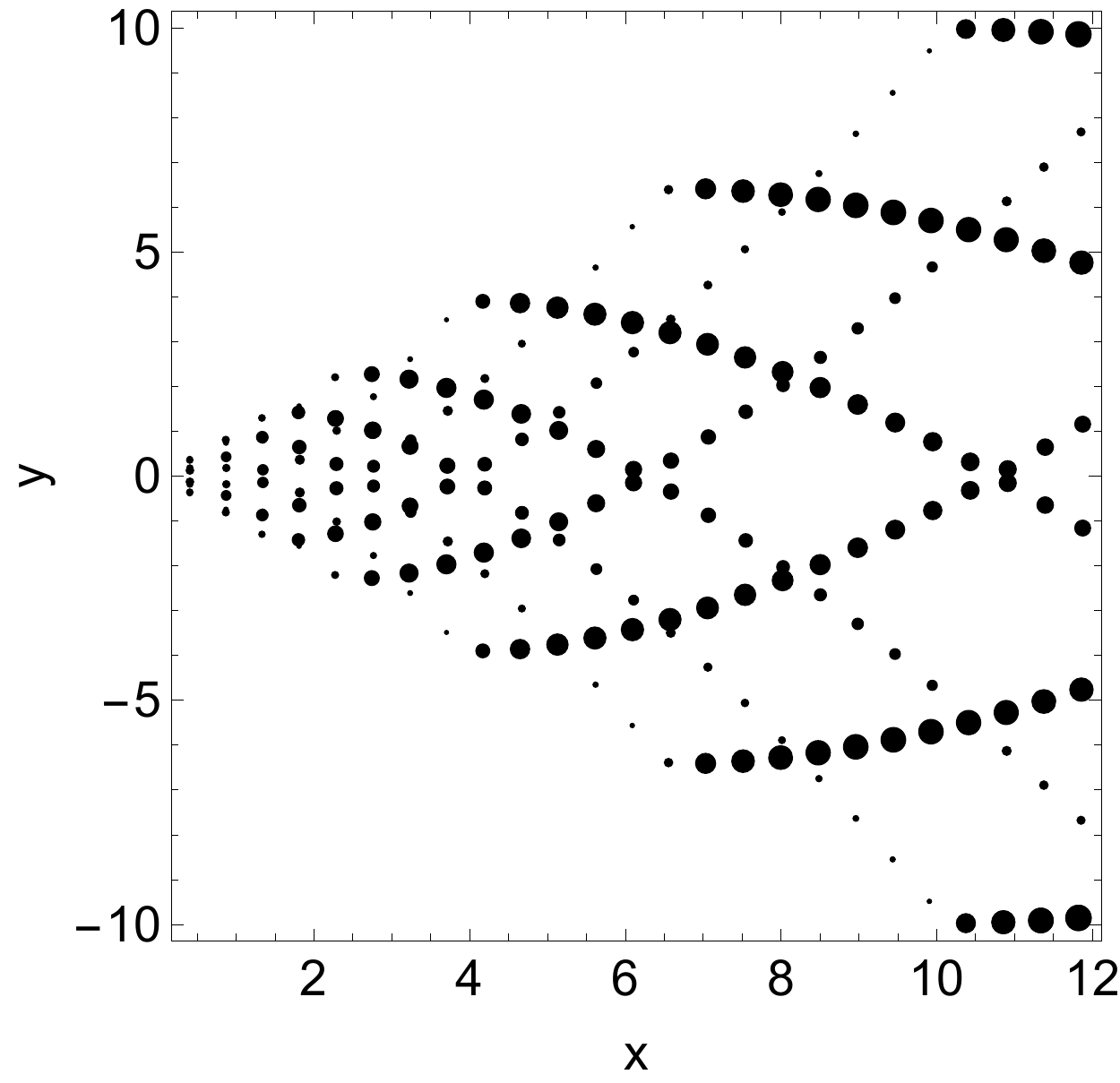}\\
\includegraphics[width=.45\columnwidth]{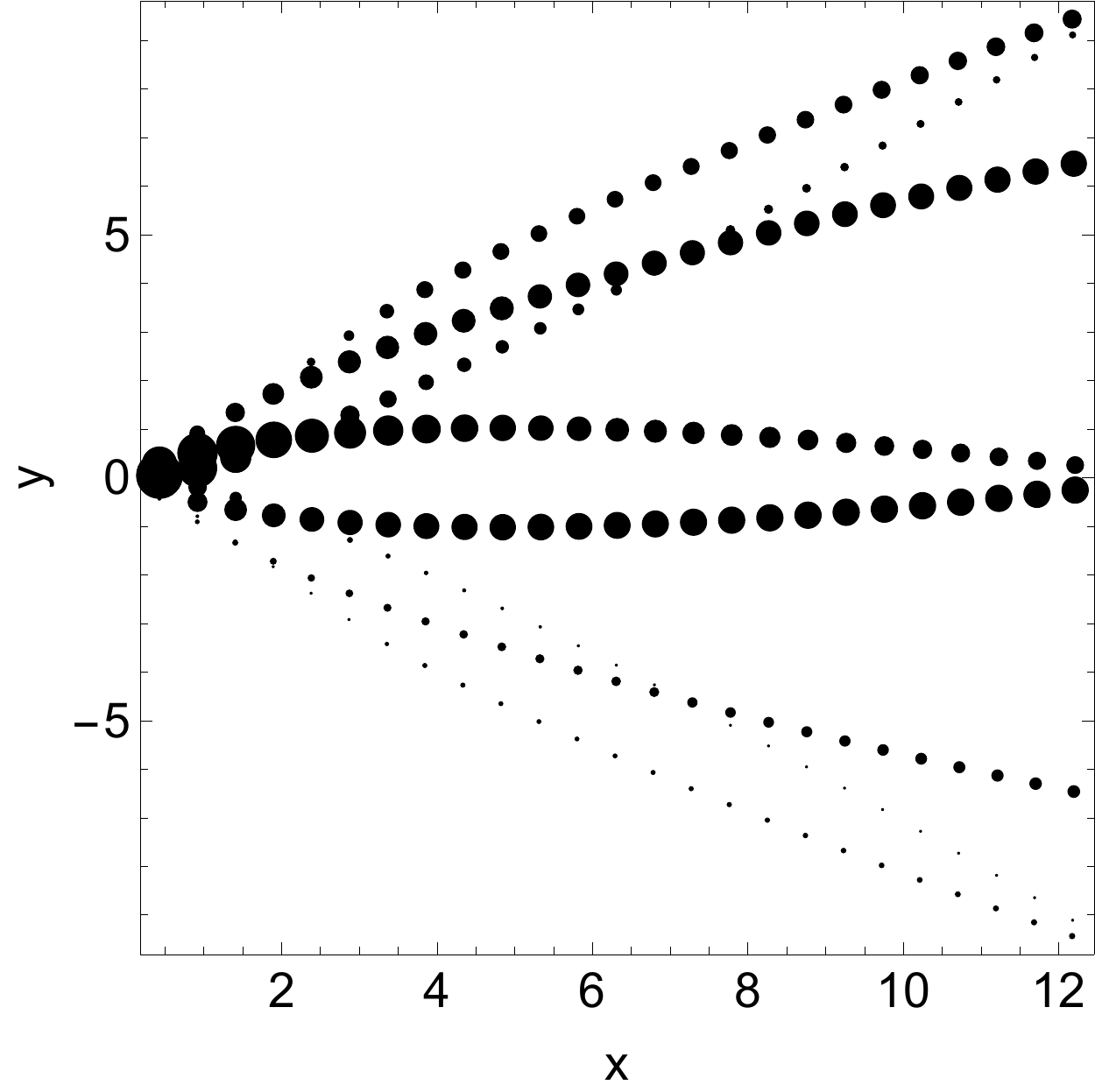}\quad
\includegraphics[width=.45\columnwidth]{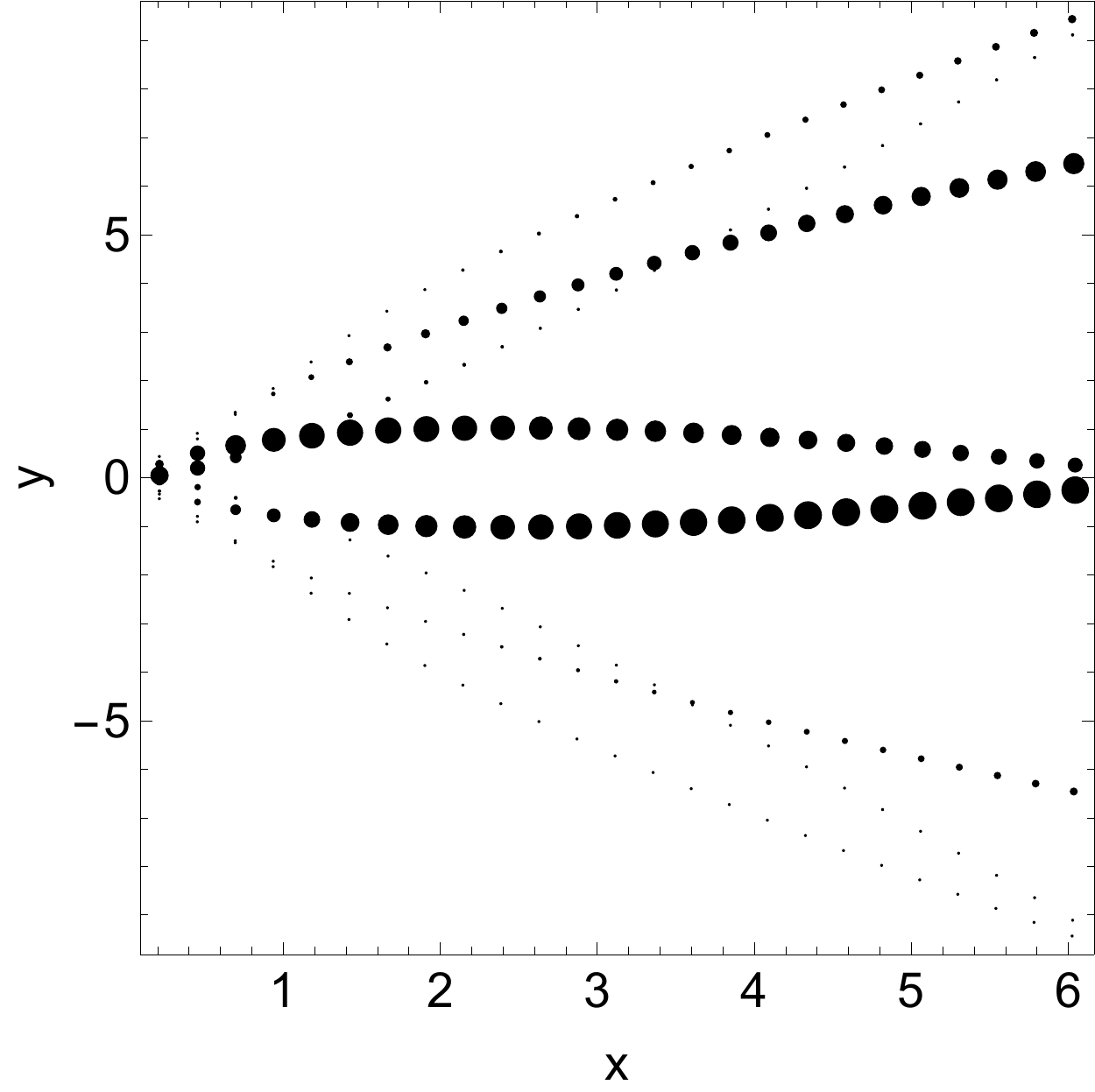}
\caption{Examples of knot trajectories. Each panel shows 8 trajectories, corresponding to launching moments separated by an eighth of a period. The size of the circles is proportional to $\delta^4$ (normalization is different in different panels).  Projection of the central axis of the jet onto the plane of the sky is along $x$ axis. 
Position of the knots are plotted for a fixed interval of the coordinate time (not observer time). 
Top row: $\theta_{ob} = 1/10$, $\phi_{ob}=0$, ${\cal  M}_j =50$, $\gamma_j =17$ (top left) and ${\cal  M}_j =10$, $\gamma_j =3.7$ (top right). 
Bottom left panel:  $\theta_{ob} = 1/10$, $\phi_{ob}=1/5$, $\gamma_j =17$, bottom right  panel:  $\theta_{ob} = 1/20$, $\phi_{ob}=1/4$, $\gamma_j =17$ ($ {\cal  M}_j $ is the Mach number of the jet in terms of the pattern velocity).} 
\label{trajectories}
\end{figure}

 We assume that the intrinsic knot emission is constant, so that the observed flux is determined exclusively by the effects of relativistic Doppler boosting; 
 which we assume $\propto \delta^4$, where
 \be
 \delta = \frac{1}{\gamma( 1- \beta \cos \theta)},
 \ee
  $\theta$ is the angle between the knot velocity and the line of sight, and $\gamma$ is the \Lf\ of the knot.  Selected knot  trajectories are  pictured in Fig. \ref{trajectories}. 

\subsection{Properties of emission patterns}

Motion of the emission feature is the combination of advection with the ballistic jet motion and propagation along the jet.  There are two qualitatively different trajectories: advection-dominated and pattern-propagation-modified regimes. In the  advection-dominated regime a given feature is mostly advected with flow, moving along a nearly ballistic trajectory. In the pattern-propagation-modified regime a feature mostly follows a curved structure of the jet.
Fig. \ref{Picture-Main} demonstrates that as the jet propagates further away from the launching site the direction of the jet becomes more and more perpendicular to the jet motion.  Thus, qualitatively, for the pattern speed with respect to the jet of the order of the speed of light, for $\theta_{max} \gamma_j \geq 1 $ a pattern is mostly advected with the flow. On the other hand for $\theta_{max} \gamma_j \leq 1 $ a pattern can propagate through a number of different wiggles.

Analyzing the figures \ref{trajectories} we come to the following conclusion: 
\begin{itemize}
\item  motion of emission knots is non-ballistic; 
\item  motion of emission knots is confined within a fixed cone; the opening angle of the cone $\theta_{open}$  is determined by the amplitude of oscillations and the viewing angle (due to small viewing angle the opening angle of knots' motion  is generally much larger than the oscillations angle, $\theta_{open} \sim \theta_{max}/ \sin \theta_{ob}$ );
\item knots may experience seemingly sudden changes of direction (top right panel);  such sudden changes are usually accompanied by brightness changes
\item  depending on the parameter 
$\theta_{max} \gamma_j$ knots experience smooth curved trajectories that seemingly asymptotically approach straight lines  (\eg\ top left panel, $\theta_{max} \gamma_j=1.7$) or oscillatory behavior 
(\eg\ top right  panel, $\theta_{max} \gamma_j=0.36$)
\item  for cases  $\theta_{max} \gamma_j \geq 1 $ typical trajectories ``bend-in'' - knots are moving toward the symmetry axis (top right); for $\theta_{max} \gamma_j \leq 1 $ some trajectories ``bend-out'' 
\item  for high Mach numbers  symmetric case (top left panel) blobs propagating along the average jet direction are brighter
\item blobs may seem to appear not at the core, with seemingly non-ballistic trajectories; knots may also seem to disappear in flux limited observations (top right and bottom left  panel)
\item At a given projected location identical knots may have very different brightness (points of intersection of  sequences  of small and large circles)
\item Motion of knots can be asymmetric with respect to the projection of the average direction of motion on the sky:  
the  brightest blobs propagate along a direction which is different from 
the jet axis (bottom rows). 
\end{itemize}

We stress that these highly variable features come from {\it constant intrinsic knot emissivity}.

\section{Discussion}

We have outlined a model  of blazar activity - a jet  carrying helical \Bf\ with a regularly changing direction. We demonstrate that this highly deterministic model
   can produce highly variable polarization,  EVPAs, and intensity profiles. At the same time the model reproduces smoothly varying EVPA changes.
Thus,  though for any given configuration the intensity, polarization and the EVPA are deterministic and thus  their behavior is highly correlated, the non-monotonic variations of these values as functions of the jet direction and \Lf\ produce highly variable overall behavior.
We find that for smooth variation of EVPA (i) $\Pi$ can be highly variable; (ii) $\Pi\sim 0$  at the moment of fastest EVPA swing; (iii) the intensity is usually maximal at points of fastest EVPA swing, but can have a minimum; (iv) for some special pitch angles there are large fluctuations of EVPA, but this always occurs at small $\Pi$. 

Importantly, these features are obtained for the assumed {\it constant} intrinsic jet emissivity. Variations in the acceleration/emissivity properties of the jets, more complicated or irregular beam trajectories, as well as existence of multiple emission components within one beam will complicate observed profiles. In addition to the possible presence of a turbulent magnetic fields, these features may produce small ($<$90$^{\circ}$), seemingly random EVPA variations \citep[\eg][]{2013ApJ...768...40L}.  Such variations are often observed during the quiescent state of the source and are not considered in this paper.


\bibliographystyle{mdpi}


\end{document}